\documentclass[structabstract]{aa}  
\usepackage{natbib}
\bibpunct{(}{)}{;}{a}{}{,} 
\usepackage{graphicx}
\usepackage{txfonts}
\begin{document}
   \title{The 2.35 year itch of Cyg~OB2\,\#9}

   \subtitle{II. Radio monitoring\thanks{Based on observations with the
              Expanded Very Large Array (EVLA), which is operated by 
              the National Radio Astronomy Observatory.
              The National Radio Astronomy Observatory is a facility of
              the National Science Foundation operated under cooperative
              agreement by Associated Universities, Inc.}
}

   \author{R. Blomme\inst{1}
          \and Y. Naz\'e\inst{2}\fnmsep\thanks{Research Associate FRS-FNRS.}
          \and D. Volpi\inst{1}
          \and M. De Becker\inst{2}
          \and R.K. Prinja\inst{3}
          \and J.M. Pittard\inst{4}
          \and E.R. Parkin\inst{5}
          \and O. Absil\inst{2}
          }

   \institute{
              Royal Observatory of Belgium, Ringlaan 3, 1180 Brussel, Belgium;
              \email{Ronny.Blomme@oma.be}
         \and
              D\'epartement AGO, Universit\'e de Li\`ege, 
              All\'ee du 6 Ao\^ut 17, B\^at. B5C, B-4000 Li\`ege,
              Belgium
         \and
              Department of Physics \& Astronomy, University College London, 
              Gower Street, London WC1E 6BT, UK
         \and
              School of Physics and Astronomy, The University of Leeds,
              Woodhouse Lane, Leeds LS2 9JT, UK
         \and
              Research School of Astronomy and Astrophysics,
              The Australian National University, Australia
             }

   \date{Received ; accepted }

 
  \abstract
  {Cyg~OB2\,\#9 is one of a small set of non-thermal radio emitting massive 
O-star binaries. The non-thermal radiation is due to synchrotron 
emission in the colliding-wind region. 
Cyg~OB2\,\#9 was only recently discovered to be a binary system and
a multi-wavelength campaign was organized to study its 2011
periastron passage.
}
   {We want to better determine the parameters
of this system and model the wind-wind collision. This will lead to
a better understanding of the Fermi mechanism that accelerates electrons up to
relativistic speeds in shocks, and its occurrence in colliding-wind
binaries. We report here on the results of the radio observations
obtained in the monitoring campaign and present a simple model
to interpret the data.}
   {We used the Expanded Very Large Array (EVLA) radio interferometer to
obtain 6 and 20~cm continuum fluxes during the Cyg~OB2\,\#9 periastron
passage in 2011. We introduce a simple model to solve the radiative
transfer in the stellar winds and the colliding-wind region, and thus
determine the expected behaviour of the radio light curve.
}
   {The observed radio light curve shows a steep drop in flux sometime before
periastron. The fluxes drop to a level that is comparable to the expected
free-free emission from the stellar winds, suggesting that the non-thermal
emitting region is completely hidden at that time. After periastron
passage, the fluxes slowly increase. We use the asymmetry of the light
curve to show that the primary has the stronger wind.
This is somewhat unexpected if we use the astrophysical parameters
based on theoretical calibrations. But it becomes entirely feasible
if we take into account that a given spectral type -- luminosity class
combination covers a range of astrophysical parameters.
The colliding-wind region also contributes to the free-free emission, which
can help to explain
the high values of the spectral index seen after periastron passage.
Combining our data with older Very Large Array (VLA) data allows
us to derive a period $P=860.0 \pm 3.7$ days for this system.
With this period, we update the orbital parameters that were derived in the 
first paper of this series.
}
   {A simple model introduced to explain only the radio data already 
allows some constraints to be put on the parameters of this binary system. 
Future, more sophisticated, modelling that will 
also include optical, X-ray and interferometric information will provide
even better constraints.}

   \keywords{stars: individual (Cyg~OB2\,\#9) -
             stars: early-type - stars: mass-loss -
             radiation mechanisms: non-thermal -
             acceleration of particles -
             radio continuum: stars}

   \maketitle
%

\section{Introduction}
\label{section introduction}

Among the early-type stars there are a number of non-thermal radio
emitters. It is now generally accepted that all such stars are 
colliding-wind binaries 
\citep{2000MNRAS.319.1005D, 2007A&ARv..14..171D, 2011BSRSL..80...67B}.
In a massive early-type binary the strong
winds from both components collide, leading to the formation
of two shocks, one on each side of the contact discontinuity
where the two winds collide. 
Around those shocks a fraction of the electrons
is accelerated up to relativistic speeds. This is believed to be due to the
Fermi acceleration mechanism \citep{1993ApJ...402..271E}.
As these electrons spiral around in the magnetic field, they emit
synchrotron radiation, which we detect as non-thermal radio emission.
In addition, the hot compressed material in the colliding-wind
region (CWR) also emits 
X-rays \citep{1992ApJ...386..265S, 2010MNRAS.403.1657P}
and influences the shape of the optical spectral
lines \citep{2005A&A...432..985R}.

While the general outline of the explanation for non-thermal
radio emission is clear, there still remain problems when 
detailed modelling of specific systems is attempted. 
In modelling the radio flux variations of the short-period variable 
Cyg~OB2\,\#8A,
\citet{2010A&A...519A.111B} failed to obtain the correct spectral index.
For the well-observed WR+O binary \object{WR~140}, models fail to 
explain the behaviour of the radio light curve 
\citep{1990MNRAS.243..662W, 1995ApJ...451..352W, 2011BSRSL..80..555P}.
Part of the problem is the difficulty in calculating all the 
effects of the Fermi acceleration ab initio. Another complication
is the presence of clumping or porosity in the stellar winds
of massive stars, making the estimates of mass-loss rates uncertain.
As an added difficulty,
the degree of clumping is dependent on the radius
\citep{2006A&A...454..625P}.

\defcitealias{PaperI}{Paper~I}
More detailed observations that will help constrain theoretical models
are therefore most important. To that purpose a multi-wavelength
campaign was started (PI: Y. Naz\'e) to monitor the 2011 periastron passage
of \object{Cyg~OB2\,\#9}. 
That Cyg~OB2\,\#9 is a binary was only discovered relatively recently.
The system consists of an O5-5.5If primary
and an O3-4III secondary in a highly eccentric ($e \approx 0.7$) orbit
with a period of $\sim$\,2.35 yr 
\citep{2008A&A...483..543N, 2008A&A...483..585V}.
Since its discovery as a binary, Cyg~OB2\,\#9 has passed 
periastron twice. The 2009 passage was unobservable because it
occurred when the system was in conjunction with the Sun.
The 2011 periastron passage is therefore the first one 
that could be well observed.

In \citet[][ hereafter \citetalias{PaperI}]{PaperI}
we presented the results from the optical and X-ray
monitoring campaign. The CWR was detected in H$\alpha$,
where it creates enhanced absorption and emission.
The X-rays, due to the hot material in the CWR,
show phase-locked behaviour with the flux peaking
at periastron. They indicate an adiabatic wind-wind collision 
for most of the time with the
flux following the predicted inverse relation with the separation
between the two components. Only close to periastron could 
the shock be turning radiative.
The present paper is the second in the series, analysing the radio observations
made by the EVLA \citep[Expanded Very Large Array, ][]{2011ApJ...739L...1P}.
Future papers will discuss the optical interferometry and the modelling
of the system.

The non-thermal nature of Cyg~OB2\,\#9 became clear from its high
brightness temperature, its non-thermal spectral index and its
variability 
\citep{1983ApJ...272L..19W, 1984ApJ...280..671A, 1989ApJ...340..518B, 1990ApJ...359L..15P}.
In a series of papers,
\citet{2004A&A...418..717V, 2005A&A...433..313V, 2006A&A...452.1011V}
tried to explain the non-thermal radio emission by a single star
(the binary nature of this star was not yet known at that time).
In a single star, it is assumed that the shocks due to the intrinsic
instability of the radiation driving mechanism are responsible
for the Fermi acceleration \citep{1985ApJ...289..698W}. 
However, the increasingly more
sophisticated models used by van Loo et al. failed to explain the observations. 

A large set of VLA archive data allowed 
\citet{2008A&A...483..585V} to find a $\sim$\,2.35 yr
period in the fluxes, strongly suggesting binarity. Simultaneously,
\citet{2008A&A...483..543N}
detected the binarity from optical spectroscopy. The orbital
information was then refined further using additional optical spectra
(\citealt{2010ApJ...719..634N}; \citetalias{PaperI}). Further evidence for the 
non-thermal nature of Cyg~OB2\,\#9
comes from VLBA (Very Large Baseline Array) radio
observations that clearly show the bow-shaped extended emission 
typical of a CWR in a binary system \citep{2006evn..confE..49D}.

In this paper, we present the EVLA (Expanded Very Large Array) radio 
observations that were 
obtained as part of the 2011 Cyg~OB2\,\#9 monitoring campaign.
We derive the orbital period from these data.
We also introduce a simple numerical model to help us interpret the 
observations.
In Sect.~\ref{section observations} we describe the data reduction
of the radio observations. 
In Sect.~\ref{section results}, we
present the radio light curve and derive the binary period. 
In Sect.~\ref{section model} we introduce a simple model, which we use
in Sect.~\ref{section discussion} to analyse and discuss the results,
Sect.~\ref{section conclusions} summarizes our
findings and presents our conclusions.

\section{Observations}
\label{section observations}

\begin{figure*}
\centering
\resizebox{\hsize}{!}{\includegraphics{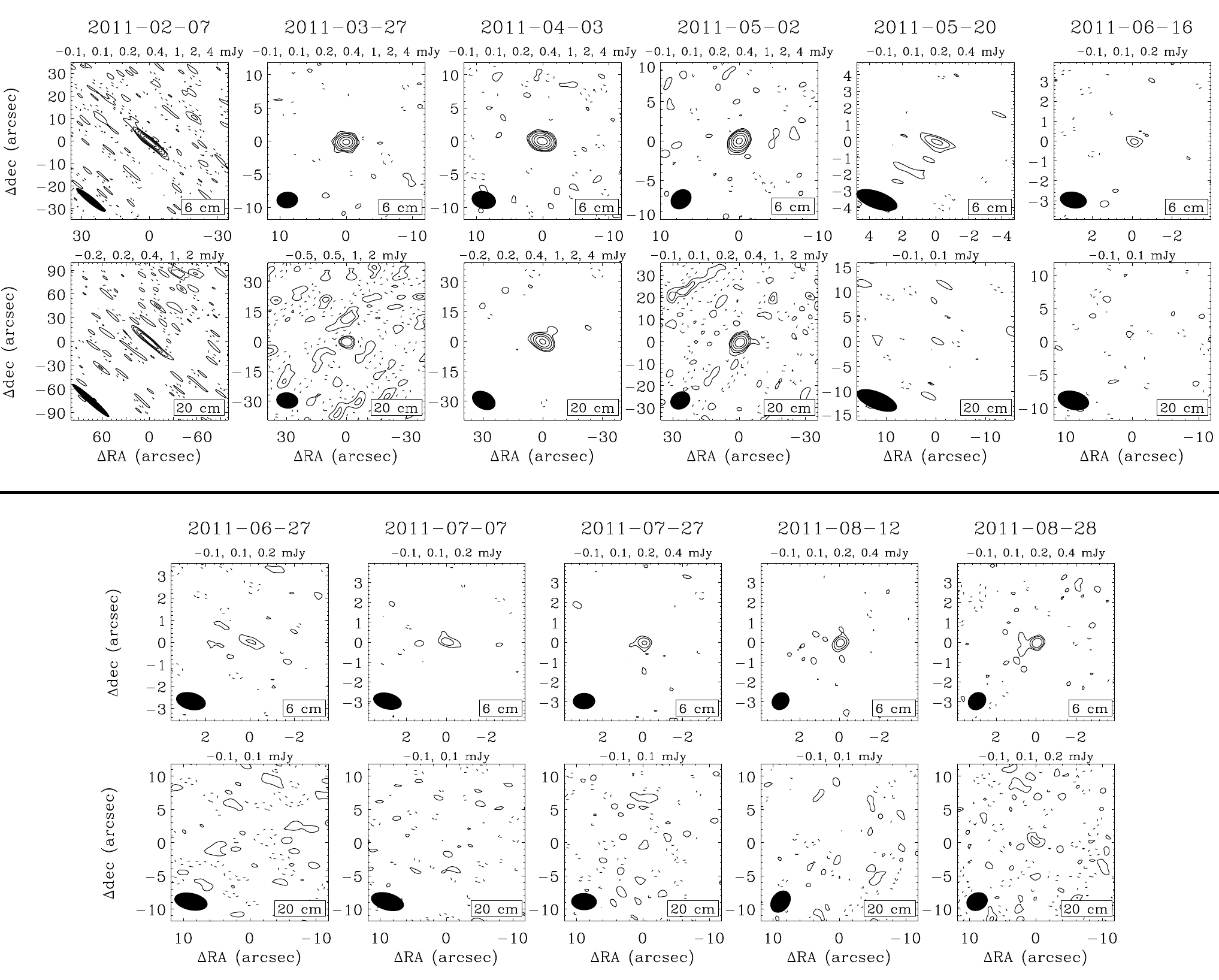}}
\caption{Radio images of Cyg~OB2\,\#9.
For each of the 11 observations, we show the C-band (6~cm) image at the
top, and the L-band image (20~cm) at the bottom. The observation date
is shown in each title. Each image shows a
small region centred on Cyg~OB2\,\#9. 
The contour levels are listed
at the top of each figure. They are shown as solid/dashed lines for
positive/negative values. The levels were chosen so that the lowest positive
level is at about $2\times$ the root-mean-square (RMS) level. The highest level is below
the peak flux value of Cyg~OB2\,\#9. The synthesized beam is shown by
the filled ellipse in the bottom left corner of each figure.
Note that the size of the image
can be different for different figures (due to the changing
configuration of the EVLA). 
}
\label{fig images}
\end{figure*}

\begin{table*}
\caption{Observing log, radio fluxes and spectral indexes
of the EVLA data on Cyg~OB2\,\#9.}
\label{table radio data}
\centering
\begin{tabular}{ccccccccccccccccccccccc}
\hline\hline
date  & HJD & orbital & config & \multicolumn{2}{c}{J2007+4029} & & \multicolumn{3}{c}{Cyg~OB2\,\#9} \\
\cline{5-6}
\cline{8-10}
      & -2\,450\,000 & phase & & C-band (6 cm) & L-band (20 cm) & & C-band (6 cm) & L-band (20 cm) & spectral \\
      &     &       & & (Jy) & (Jy) & & (mJy) & (mJy) & index \\
2011-02-07 &    5599.51 & 0.836 & CnB & 3.540 $\pm$ 0.017 & 2.158 $\pm$ 0.067 &  &  6.91 $\pm$  0.06 &  4.01 $\pm$  0.12 &  0.43 $\pm$  0.02 &  \\
2011-03-27 &    5648.28 & 0.892 &   B & 3.541 $\pm$ 0.004 & 2.088 $\pm$ 0.024 &  &  5.60 $\pm$  0.04 &  3.89 $\pm$  0.32 &  0.29 $\pm$  0.07 &  \\
2011-04-03 &    5655.28 & 0.901 &   B & 4.237 $\pm$ 0.029 & 2.077 $\pm$ 0.022 &  &  6.68 $\pm$  0.05 &  5.01 $\pm$  0.07 &  0.23 $\pm$  0.01 &  \\
2011-05-02 &    5684.16 & 0.934 &   B & 3.675 $\pm$ 0.003 & 2.159 $\pm$ 0.040 &  &  5.26 $\pm$  0.05 &  3.91 $\pm$  0.06 &  0.23 $\pm$  0.01 &  \\
2011-05-20 &    5702.09 & 0.955 & BnA & 3.592 $\pm$ 0.006 & 2.057 $\pm$ 0.013 &  &  0.49 $\pm$  0.05 & $\le$ 0.16 & $\ge$ 0.91 &  \\
2011-06-16 &    5729.12 & 0.986 &   A & 3.278 $\pm$ 0.005 & 1.893 $\pm$ 0.012 &  &  0.27 $\pm$  0.04 & $\le$ 0.15 & $\ge$ 0.48 &  \\
2011-06-27 &    5740.07 & 0.999 &   A & 3.278 $\pm$ 0.005 & 1.948 $\pm$ 0.009 &  &  0.27 $\pm$  0.05 & $\le$ 0.19 & $\ge$ 0.26 &  \\
2011-07-07 &    5750.06 & 0.011 &   A & 3.240 $\pm$ 0.005 & 1.901 $\pm$ 0.013 &  &  0.35 $\pm$  0.04 & $\le$ 0.16 & $\ge$ 0.60 &  \\
2011-07-27 &    5769.94 & 0.034 &   A & 3.371 $\pm$ 0.005 & 1.971 $\pm$ 0.005 &  &  0.49 $\pm$  0.04 & $\le$ 0.17 & $\ge$ 0.81 &  \\
2011-08-12 &    5785.86 & 0.052 &   A & 3.443 $\pm$ 0.005 & 2.013 $\pm$ 0.006 &  &  0.72 $\pm$  0.04 & $\le$ 0.17 & $\ge$ 1.13 &  \\
2011-08-28 &    5801.81 & 0.071 &   A & 3.361 $\pm$ 0.006 & 2.008 $\pm$ 0.008 &  &  0.89 $\pm$  0.05 &  0.29 $\pm$  0.06 &  0.88 $\pm$  0.17 &  \\
\hline
\end{tabular}
\tablefoot{We list observing date, heliocentric Julian date, orbital phase
(using the Sect.~\ref{section period} value for the period)
and configuration of the EVLA. The fluxes of the phase calibrator 
J2007+4029 are listed next (at 6 and 20~cm), followed by the 6 and 20~cm 
fluxes of Cyg~OB2\,\#9 and their spectral index.
All fluxes have been calibrated on the flux calibrator 3C147.
}
\end{table*}

\subsection{EVLA data}

Cyg~OB2\,\#9 was monitored with the EVLA during 
the period January to August 2011. The observations (programme 10C-134) were 
obtained through the Open Shared Risk Observing (OSRO) programme. 
Eleven observations were made, each time in C-band (4.832 -- 5.086 GHz, 6~cm) 
and L-band (1.264 -- 1.518 GHz, 20~cm). 
The observing log is given in Table~\ref{table radio data}.
During the monitoring period, the configuration of the EVLA changed 
from CnB (giving a lower spatial resolution) to A 
(the highest spatial resolution).

Each band is covered by 2 $\times$ 64 channels, 
each of 2 MHz width. To calibrate out instrumental and atmospheric effects,
the phase calibrator \object{J2007+4029} is used.
An observation consists of a phase calibrator -- target
-- phase calibrator sequence, first in C-band, then in L-band. This sequence
is followed by an observation of the flux calibrator (J0542+498=\object{3C147})
in L and C band. The flux calibrator also serves as the bandpass calibrator.
Time on target for a single observation
is 5 min for C-band and 7 min for L-band.

\subsection{Data Reduction}
\label{section data reduction}

The data were reduced using the 
CASA\footnote{{\tt http://casa.nrao.edu/}}
(Common Astronomy Software Applications)
version 3.3.0 data reduction package.
The system already flags a number of problematic data (due to focus problems,
incorrect subreflector position, off-source antenna position or missing antennas)
and other flags are applied while reading in the data (shadowing by antennas).
Careful attention
is given to Radio Frequency Interference (RFI). The small amount of RFI
present in the C-band is removed by flagging the visibilities
in the relevant channels.
For the stronger and more extended RFI in L-band, we apply Hanning smoothing
to remove the Gibbs ringing; after that the channels affected by the
RFI are flagged.

The calibration sequence starts by assigning the correct flux to
the flux calibrator (3C147), using a model to take into account that this
source is slightly resolved.
A preliminary gain phase
calibration is applied before the delay and bandpass calibration.
This is followed by the final gain phase and gain amplitude calibrations.
The fluxscale is then transferred from the flux to the phase calibrator
(J2007+4029).
The phase calibrator fluxes are listed in Table~\ref{table radio data}:
they show the slow flux variations typical of many phase calibrators.
The calibrations are then applied to the flux and phase calibrators,
as well as to the target. The calibrated data for flux and phase calibrators
are inspected visually: if discrepant data are found, they are flagged and the
calibration is re-done. 

In L-band (20~cm), the phase calibrator is influenced by the radio galaxy
\object{Cyg A}, 
which contributes about 0.05 -- 0.5 Jy (depending on the configuration
of the EVLA) to the image, while the phase
calibrator itself is about 3 Jy. As the sidelobes of such a strong source
can influence the phase calibrator, we apply self-calibration
to improve the gain phases.
The visibilities are then inspected and any discrepant
data are flagged. 

In the next step 
an image is made from the target visibilities. For the C-band, this image
covers an area somewhat
larger than the primary beam (which is 17\arcmin\ diameter).
To gain computing time, we make the image in the L-band somewhat smaller
than the primary beam (which is 60\arcmin\ diameter).
The size of each (square) pixel is chosen such that it oversamples
the synthesized beam by a factor of at least 4 in each dimension.
The image is cleaned down to a level where the noise in the centre
is compatible with the expected noise.
Images taken in the low-spatial resolution EVLA configurations contain
substantial extended Galactic background. This is removed by excluding
the data on the shortest baselines during the imaging.
A strong advantage of the present EVLA data over older VLA continuum data
is that the many channels allow a sharp image further away
from the field centre. To avoid introducing artefacts in such wide-field
imaging \citep{2008A&A...487..419B},
we need to use the specific wide-field options in the CASA cleaning procedure.
To handle the smaller-scale extended emission still present in the
L-band images, the multi-scale option in the CASA clean procedure
is used, with scales chosen to be 5, 10, 20 and 40 times the pixel size.
For those images where the Cyg~OB2\,\#9 flux is high enough 
($\ga 1$~mJy), we also apply a single round of phase-only 
self-calibration (further rounds of self-calibration no longer 
improve the image).

Two of our observations in the L-band (20~cm) are strongly affected by flaring
from \object{Cyg X-3} (which is within the primary beam). 
This microquasar consists of a Wolf-Rayet star
and a compact object (most likely a black hole), and
shows frequent and strong
flaring activity. The Cyg X-3 multi-wavelength monitoring campaign 
reported by \citet{2012MNRAS.421.2947C} shows a sharp transition
from a quenched state to a major flare, with an onset estimated
at MJD $55641.0 \pm 0.5$ (i.e. March 21).
On our March~27 observation, Cyg~X-3 has a flux of $\sim$\,10~Jy; on
the April~03 observation this has decreased to $\sim$\,1~Jy, and on the
May 02 observation it has dropped further down to
$\sim$\,0.07~Jy. 
This behaviour is compatible with the \citeauthor{2012MNRAS.421.2947C}
monitoring results.
Because of the decreasing sensitivity away from the
field centre, the contribution of Cyg~X-3 to our observations
is a factor $\sim$\,3 less
than the numbers given above.
Nevertheless, for the March~27 and April~03 observations,
the sidelobes of Cyg~X-3 strongly
perturb the cleaning of the image and the measurement of Cyg~OB2\,\#9,
which has a flux that is only a few mJy at best.
The effect of the Cyg~X-3 sidelobes
on earlier and later observations is negligible.

For the two observations most affected, we therefore first
make an image, limiting the clean components to a small box around
Cyg~X-3. We use multi-frequency synthesis, resulting in
two images, one with the flux (averaged over the frequency band) and
another with the spectral index. We then self-calibrate these images
using one step of phase-only calibration, followed by one step
of amplitude and phase calibration. The clean components of the Cyg~X-3 image
are then subtracted from the visibility data of the target. These
data are then processed in the standard way to make an image and measure
the fluxes. The procedure is very successful for the April~03
observation, but in the March~27 observation important residual effects
of Cyg~X-3 remain. In cleaning the latter image, we therefore do not
attain the expected noise level. We do not apply self-calibration
to this image either.

For all images resulting from our dataset, we then
determine the flux and its corresponding error bar
by fitting an elliptical Gaussian to the target. 
To measure the Cyg~OB2\,\#9 flux, we fix the size and position angle of the
beam to the values of the synthesized beam
(which is the shape that a point source should have after cleaning the
image).
In a number of L-band (20~cm) observations,
Cyg~OB2\,\#9 is not detected. In such cases, we assign an upper limit of
three times the root-mean-square (RMS) noise measured around the target
position.

From the fluxes at 6 and 20~cm, we derive the spectral index $\alpha$,
given by $F_\nu \propto \nu^{\alpha}$. The error bar on $\alpha$ is
derived from standard error propagation, using the error bars on each
of the fluxes. Where the 20~cm flux has an upper limit, only a lower
limit for $\alpha$ can be determined. 

\section{Results}
\label{section results}

\subsection{Radio light curve}
\label{section radio light curve}

\begin{figure}
\centering
\resizebox{8.5cm}{!}{\includegraphics{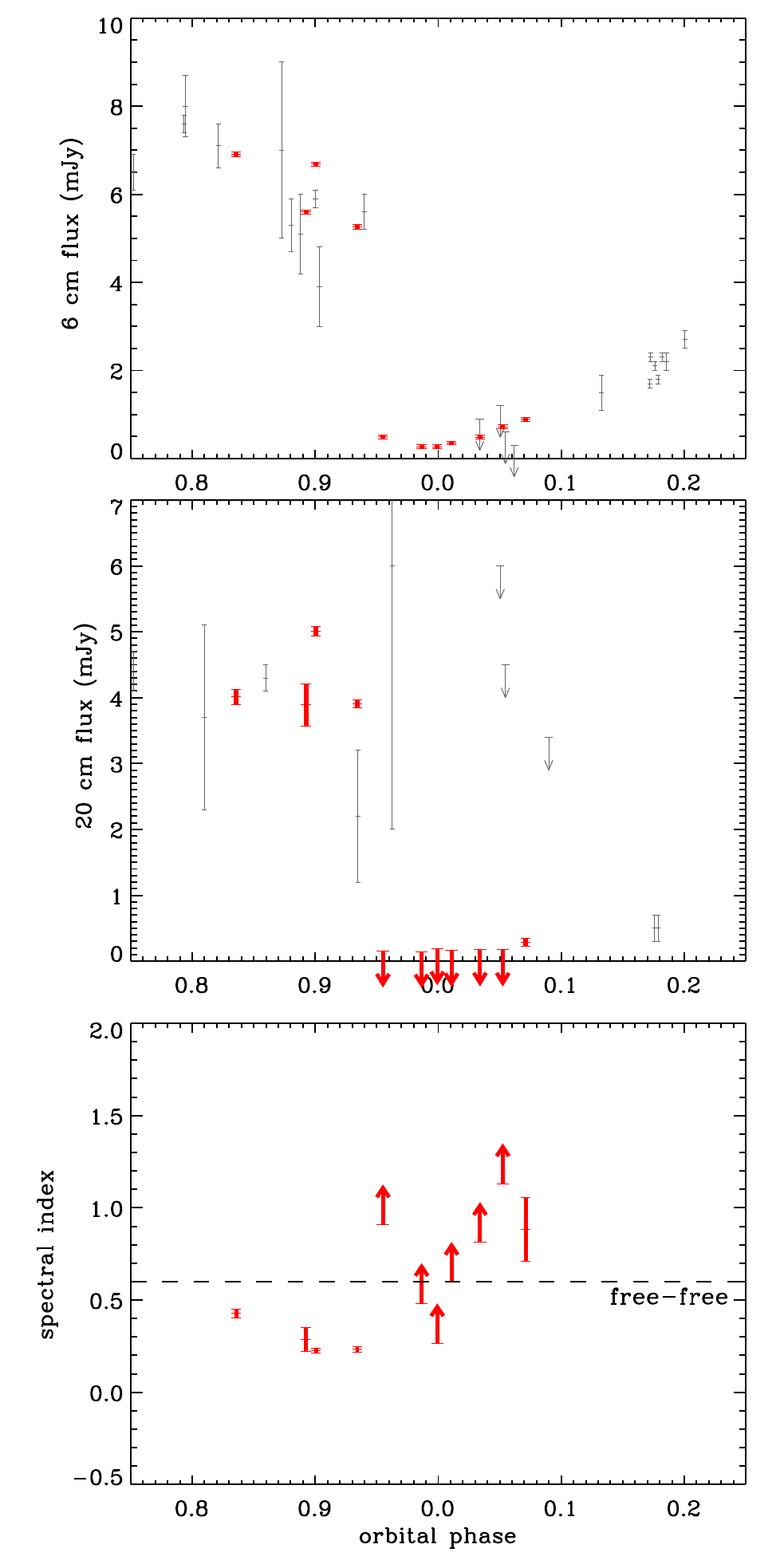}}
\caption{Radio fluxes and spectral indexes around the periastron passage
of Cyg~OB2\,\#9, as a function of orbital phase. The top figure shows
the 6~cm flux, the middle one the 20~cm flux and the bottom one
the spectral index.
The older VLA data from \citet{2008A&A...483..585V}
are plotted with a grey line. The new EVLA data are shown with the
thicker solid red line. The length of the line indicates the $1 \sigma$
error bar on the flux. For the non-detections, upper limits are shown that are
$3\times$ the RMS noise in the centre of the image. Note that the fluxes
have an additional uncertainty of $\sim$~5\% due to the absolute flux
calibration (not shown on the figure).
On the spectral index figure, the dashed line indicates the
value for free-free emission ($\alpha = +0.6$).
Periastron occurs at phase=0.0.
}
\label{fig lightcurve}
\end{figure}

Fig.~\ref{fig images} shows contour plots of a small region around 
Cyg~OB2\,\#9 for each of the 11 observations, at both 6 and 20~cm.
The 6~cm images show a clear decrease of the flux with time, with a
slight increase again towards the end of the series. The 20~cm series
mirrors that behaviour, with the star being undetectable from
May~20 to August~12. The changing sizes of the synthesized beam
reflect the changes of the EVLA antenna configuration.
Although we know that Cyg~OB2\,\#9 is a colliding-wind binary,
the EVLA data do not allow us to resolve the CWR.
Our resolution is at best $\sim$\,0.5\arcsec~(at 6~cm), but the VLBA
data of \citet{2006evn..confE..49D} show that the CWR has
a size of $\sim$\,0.015\arcsec\ (at 3.6~cm).

The March~27 20~cm observation is of lesser quality.
This is due to the strong influence of Cyg~X-3 on these data
(see Sect.~\ref{section data reduction}). Our data reduction procedure 
manages to remove the largest effects of Cyg~X-3 but the quality of the data
does not allow us to make an image with a sufficiently high dynamic range.

The measured Cyg~OB2\,\#9 fluxes, their error bars and the spectral
indexes are listed in Table~\ref{table radio data}. Fig.~\ref{fig lightcurve}
plots them as a function of orbital phase, with phase=0.0 corresponding
to periastron. 
The epoch of periastron passage and the period were taken from
this paper (see Sect.~\ref{section period}).
The comparison with the
older 6~cm VLA data from \citet{2008A&A...483..585V} shows good agreement,
showing that the behaviour of Cyg~OB2\,\#9 repeats very well over a number
of orbital cycles. Although each of our data points is based on only
5 -- 7 min on-target time, the quality of these EVLA data is clearly
much better than that of the older observations.
For the values in the high-flux regime, the main uncertainty is due
to the absolute flux calibration (estimated to be at the 
$\sim$~5\% level), not to the noise.
For the 20~cm data the agreement is also good, though less constraining
because of the large error bars and high upper limits of
the older VLA data set.

The new data show a strong drop in flux (both at 6 and 20~cm)
between the May 02 and May 20 observation, corresponding
to phase 0.934 -- 0.955. Before that time
the 6~cm fluxes were in a high-flux regime, but with the flux
slowly decreasing. 
The corresponding
20~cm fluxes are relatively constant during that time.
The slight increase between the March~27 and April~03 flux
at 20 cm
(phase 0.892 -- 0.901) could be due to problems with
the data reduction (see Sect.~\ref{section data reduction})
but this is less likely for the 6~cm increase. 

After the drop in flux, there is still a slight decrease in the 6~cm flux,
just up to periastron passage. After periastron the flux starts to slowly
increase again. The older VLA data at later phases connect very well
with the new data points, also in this low-flux regime. 
At 20~cm the fluxes are so low that Cyg~OB2\,\#9
is not detected. Only the last data point we have (August 28, phase 0.071)
shows a detectable but low flux. 
Again the new data connect well with the increasing 20~cm flux
shown in the older VLA data.

Previously, little information was available about the spectral
index ($\alpha$) and its changes in Cyg~OB2\,\#9 \citep{2008A&A...483..585V}.
The present data provide interesting information of
the important changes of $\alpha$ during periastron
passage (Fig.~\ref{fig lightcurve}, bottom panel). Before periastron,
$\alpha$ goes down from 0.43 
to 0.23. As this is well below
the $+0.6$ value expected for free-free emission, it clearly shows
the presence of a non-thermal component. From the May 20 (phase 0.955)
observation onwards, the index increases; at times even the lower limit
indicates a value for free-free emission for a spherically symmetric wind,
or an even higher value.

One would expect 
the synchrotron emission from the relativistic electrons in the CWR
to reach its maximum
close to periastron, where the stellar separation is smallest
and the local magnetic field strongest.
However, the observed situation here is almost the reverse.
Both the drop in flux and the rise of spectral
index point to a significant impact of free-free absorption on
the synchrotron emission component by the stellar wind material close
to periastron passage. This fact emphasizes the importance of
orientation effects in the light curve of colliding-wind binaries,
notably in the radio domain.

\subsection{Period}
\label{section period}

The new EVLA data, in combination with the older 
VLA data \citep{2008A&A...483..585V} cover a 
time span of $\sim$\,30 years, which is substantially larger than
the time span covered by the optical spectroscopy (Paper I).
The radio data are thus more suitable to derive the orbital period of
Cyg~OB2\,\#9.

As in \citet{2008A&A...483..585V}, we use the string-length method
\citep{1983MNRAS.203..917D} to determine the period that best fits the
data. We limit ourselves to the 6~cm observations, as the coverage at other
wavelengths is sparser, and we exclude upper limits. 
We normalize the fluxes to the maximum
flux over the combined VLA and EVLA dataset. This maximum is 8.5 mJy 
(VLA observation on 1984 November 27). The flux normalization
introduces a good balance between phase difference and
flux difference in the string-length calculation. 

Intrinsically, the string-length method does not provide an error
bar on its result. To determine the error bar, we apply the
bootstrap technique. We make 5\,000 Monte-Carlo simulations, where
we randomly choose a set of $n$ data points
out of the existing $n$ observations (with
replacement). We then apply the above string-length method, each time exploring
10\,000 periods between 750 and 950 days.
We thus end up with a set of 5\,000 period determinations
from which we can derive the best value and the error bar. For the best
value, we use the median and for the error bar we use quantiles to select
the middle range that contains 68.3\% of the values
(this corresponds to $\pm 1 \sigma$ for a Gaussian distribution).
We find $P=860.0 \pm 3.7$ days. Within the error bar, this is the
same result as listed in \citetalias{PaperI} (that value was based
on a preliminary reduction of the radio data).

With this new value for the period, we re-determine the orbital
solution based on the optical spectra from \citetalias{PaperI}.
The results are presented in Table~\ref{table orbital solution}.
All phases presented in the present paper are based on this
solution.

\begin{table}
\caption{Orbital solution for Cyg~OB2\,\#9.}
\label{table orbital solution}
\begin{center}
\begin{tabular}{lc}
\hline\hline
Parameter & Value \\
\hline
$P\,(d)$ - this paper     & $860.0         \pm   3.7$ \\
$T_0$                   & $4020.72       \pm   2.55$ \\
$e$                     & $0.710         \pm   0.016$ \\
$\omega_1\,(\degr)$       & $191.9         \pm   2.9$ \\
$M_1/M_2$               & $  1.13        \pm   0.08$ \\
$\gamma_1$\,(km s$^{-1}$) & $-33.9         \pm   2.7$ \\
$\gamma_2$\,(km s$^{-1}$) & $  1.0         \pm   2.8$ \\
$K_1$\,(km s$^{-1}$)      & $ 61.1         \pm   3.0$ \\
$K_2$\,(km s$^{-1}$)      & $ 68.9         \pm   3.4$ \\
$a_1 \sin i\, (R_{\sun})$ & $730.1         \pm  39.8$ \\
$a_2 \sin i\, (R_{\sun})$ & $823.2         \pm  44.9$ \\
\hline
\end{tabular}
\end{center}
\tablefoot{Based on the period derived in Sect.~\ref{section period}
and the optical spectra from \citetalias{PaperI}.
The present table supersedes the values given in \citetalias{PaperI}.
$T_0$ corresponds to periastron passage, in HJD$- 2\,450\,000$.}
\end{table}

We can also check if the epoch of periastron passage detected in the radio
corresponds to that in the optical spectroscopy
\citepalias{PaperI}.
Naively, one may expect minimum radio flux to occur at 
periastron passage: at that time the stars are at their closest
approach and any non-thermal emission from the CWR
will be largely, or totally, absorbed by the stellar wind
material.
The substantial drop in 6 and 20~cm flux occurs between phase
0.934 and 0.955, i.e. about 1 -- 2 months before periastron
passage. The lowest 6~cm flux occurs around phase 0.986 -- 0.999.
Fitting a parabola through the six lowest 6~cm fluxes gives
the minimum at 0.994. There is therefore a small offset between
periastron passage and minimum radio flux.

\begin{figure}
\centering
\resizebox{9cm}{!}{\includegraphics[bb=24 32 275 322]{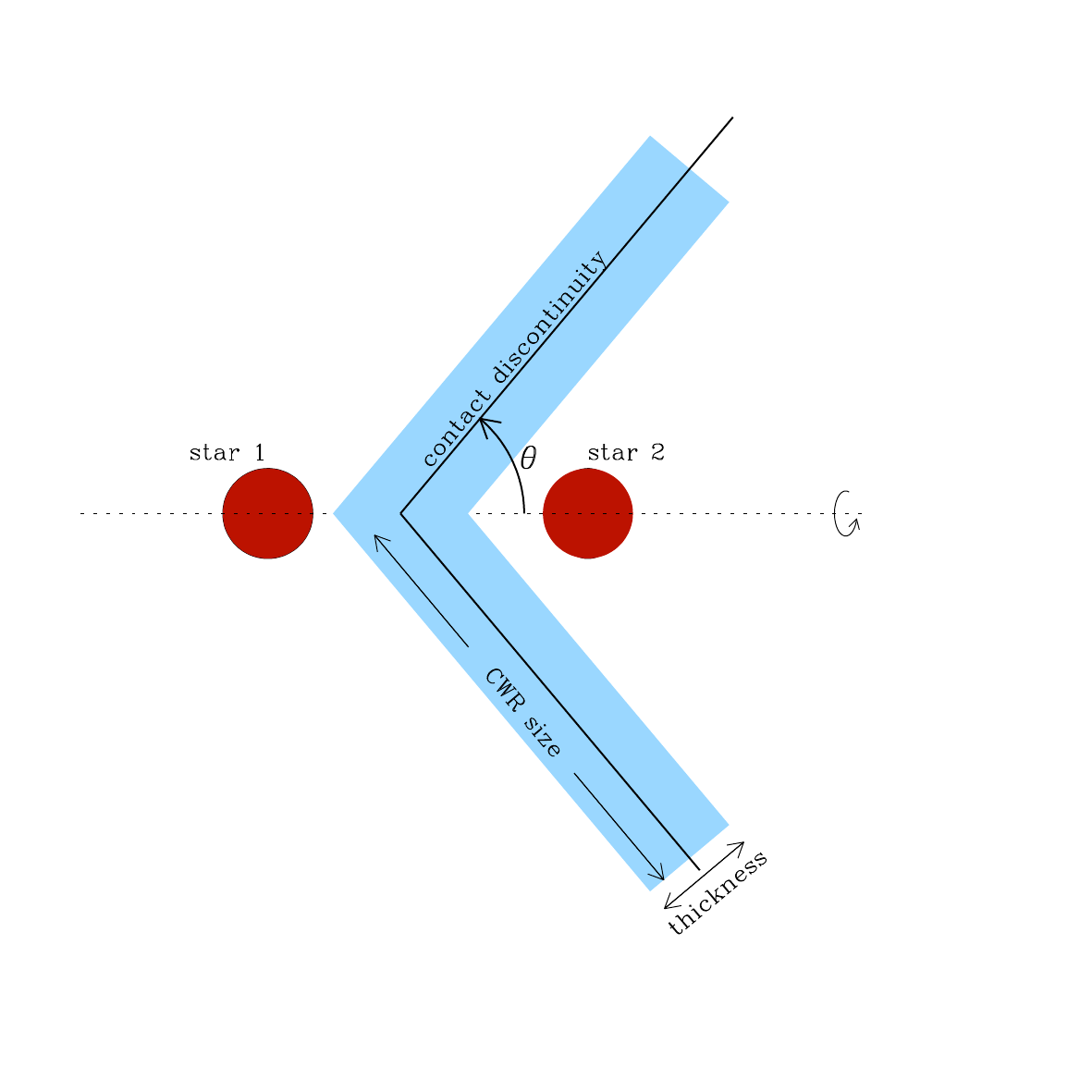}}
\caption{Schematic view of our model for the CWR. The shape of the CWR
(shaded in light-blue)
is a cone that is rotationally symmetric around the axis connecting the
two stars. It has a (half) opening angle $\theta$.}
\label{fig CWR}
\end{figure}

\section{Model}
\label{section model}

For the further analysis of the radio data it will be interesting
to have a simple model. This model should be capable of handling
the free-free emission from the stellar winds and the CWR, as well
as the non-thermal synchrotron emission from the CWR.

In the model we solve the radiative transfer equation in 
a three-dimensional grid using an adaptive grid scheme.
Our first-level grid has
$128^3$ cells, with the origin at the centre of mass
of the binary
and the negative z-direction towards the observer.
While solving the radiative transfer equation, we refine this grid
locally, as needed, to achieve the pre-specified precision on
the specific intensity.
The grid extends 12\,000~$R_{\sun}$ on either side of the origin.
At any phase 
in the orbit, we position the two stars in our 3D grid.
We take into account the estimated 62\degr\ inclination angle
\citepalias{PaperI}.
We next assume a mass-loss rate and terminal velocity for
both components.
We can then calculate the position of the collision
along the line connecting the two stars,
as well as the opening angle of the CWR
\citep[][their Eqs. (1) and (3)]{1993ApJ...402..271E}.
We simplify the shape of the contact discontinuity
by assuming it to be a cone which is
rotationally symmetric around the axis connecting the two stars
(Fig.~\ref{fig CWR}).
The size of the CWR is limited to a radius that is 
proportional to
the separation between the two components.
We also assign a thickness to the CWR. 

At any given point in our grid, the mass density can then
be determined from the mass-loss rate and terminal velocity
of the relevant star. Within
the assumed thickness of the CWR, we increase this density by a factor
4 to account for the (presumed strong) shock the material has gone through. 
Note that we ignore any possible clumping or porosity in the
wind material.
We also assign a temperature to
each point. For the unperturbed stellar wind material this 
value ($T_{\rm wind}$ = 20\,000~K) is about half
the effective temperature of the star. For the material in the CWR
we assign a temperature ($T_{\rm CWR}$) appropriate for the heated material.
This temperature is assumed to be constant over the whole CWR.
It is also assumed to be independent of 
orbital phase.
This is mainly motivated by the fact that the CWR temperature 
is related to the
pre-shock velocity, and that this velocity is not expected to change
significantly along the orbit as the winds will have reached
their terminal velocity (except close to periastron).
We then solve the radiative transfer equation following the procedure 
outlined in \citet{1975MNRAS.170...41W}. 
The radiative transfer takes into account the free-free absorption and
emission, both from the CWR and the stellar wind material.

To include the synchrotron emission in the above model,
we manipulate the temperature
and opacity we assign to the material.
As before, for
the unperturbed stellar wind material we assign a temperature that is 
about half
the effective temperature of the star. For the material in the CWR
we assign two temperatures. 
One represents
the hot, non-relativistic material in the CWR. The other 
temperature is to be interpreted as a brightness temperature,
representing the relativistic electrons that are responsible for
the synchrotron emission. In this way,
we avoid detailed and complicated calculations needed to determine
the exact synchrotron emission. 
All material emits at either the wind temperature, or the combined
CWR and synchrotron brightness temperature (if it is in the CWR).
For the opacity, the material absorbs at either wind temperature
or CWR temperature (the synchrotron brightness temperature does not
play a role in absorption).
In this way the synchrotron emission can be
absorbed by the stellar wind material as well as by the 
hot, non-relativistic material in the CWR.
We then solve the radiative transfer equation using the adaptive grid
scheme, and determine
the flux at a number of orbital phases.

\section{Analysis and discussion}
\label{section discussion}

\subsection{Free-free contribution stellar winds}
\label{section winds}

\begin{table}
\caption{Star and wind parameters of Cyg~OB2\,\#9.}
\label{table parameters}
\centering
\begin{tabular}{lcc}
\hline\hline
 & Primary & Secondary \\
Spectral type                                &    O5 -- O5.5I &    O3 -- O4III \\
$T_{\rm eff}$ (K)                            & 38520 -- 37070 & 42942 -- 41486 \\
log $L_{\rm bol}/L_{\sun}$                   &  5.87 -- 5.82  &  5.92 -- 5.82 \\
$M_*$ ($M_{\sun}$)                           & 50.87 -- 48.29 & 58.62 -- 48.80 \\
$\varv_\infty$~(km s$^{-1}$)                 &  2079 -- 2041  &  2436 -- 2303 \\
$\dot{M}$ ($10^{-6}~M_{\sun}\,{\rm yr}^{-1}$) &  5.66 -- 4.45  &  6.58 -- 4.98 \\
6 cm flux (mJy)                              &  0.25 -- 0.18  &  0.25 -- 0.18 \\
20 cm flux (mJy)                             &  0.12 -- 0.09  &  0.12 -- 0.09 \\
Radius ($R_{\sun}$) where $\tau_{\rm 6~cm}=1$  & 1712 -- 1502   & 1621 -- 1420 \\
Radius ($R_{\sun}$) where $\tau_{\rm 20~cm}=1$ & 3964 -- 3479   & 3754 -- 3287 \\
\hline
\end{tabular}
\tablefoot{For each parameter, the range corresponds to the range
in spectral types. The stellar parameters are from the calibration of 
\citet{2005A&A...436.1049M}, the
wind parameters are derived from the \citet{2001A&A...369..574V} equations
and the radio fluxes from the \citet{1975MNRAS.170...41W}
equation. A distance of 1.45~kpc is assumed.
}
\end{table}

Near minimum the spectral index is close to thermal, suggesting that the
non-thermal contribution has dropped to zero. The free-free emission
and absorption in the stellar winds of course provide a thermal component
and we now check if these low fluxes can be explained by the free-free
emission of the winds only.

In Paper I, we took the stellar parameters from \citet{2005A&A...436.1049M}
and assigned the supergiant values 
of effective temperature ($T_{\rm eff}$), luminosity 
($L_{\rm bol}$) and mass ($M_*$)
to both primary and secondary.
From the equations of \citet{2001A&A...369..574V} we then derive 
the mass-loss rate ($\dot{M}$) and the terminal velocity
($\varv_{\infty}$).
We can then use the equations of
\citet{1975MNRAS.170...41W} to calculate the
expected radio fluxes of a single-star wind at 6 and 20~cm.
Combining these results for both components
gives a 6~cm free-free flux of 
$\sim$\,1 mJy, and a 20~cm flux of $\sim$\,0.5~mJy. 
This is much too high
compared to the observed 6~cm minimum of $0.27 \pm 0.04$~mJy
and the 20~cm upper limit of 0.15~mJy.

For the purposes of this paper we therefore need more refined values
for the stellar and wind parameters. Instead of a single value,
we will consider the range
of values corresponding to the range of spectral types, both
for primary and secondary.
In Table~\ref{table parameters} we give the star and wind parameters
used here, based on the calibration by 
\citet[][theoretical $T_{\rm eff}$ scale]{2005A&A...436.1049M}.
As in \citetalias{PaperI} we use the \citet{2001A&A...369..574V} and
\citet{1975MNRAS.170...41W} equations to
derive the mass-loss rate, terminal velocity and expected radio fluxes,
as well as the radius where the radial optical depth equals 1.
The flux contributions of both stars turn out to be about equal
(the radio flux depends on the combination $\dot{M}/\varv_{\infty}$,
which is about the same for these stars).
The sum of both fluxes is still higher than the observed 6~cm minimum
flux of $0.27 \pm 0.04$~mJy. 
The fluxes are lower however than the \citetalias{PaperI} fluxes
because here we used the \citet{2005A&A...436.1049M} giant values instead
of supergiant values for the secondary.

A number of effects can lessen the discrepancy between the predicted
fluxes and the observed minimum.
First of all, the total flux is not given by the straightforward sum
of both fluxes. The winds of both stars collide at the contact 
discontinuity; on each side of this discontinuity there is only
material belonging to one star, not to the other. The contact 
discontinuity falls well within the radii of optical depth~=~1
(Table~\ref{table parameters}). This suggests that
a substantial part of the summed flux will be missing.
(A more correct description takes into account the
material accumulated between the contact discontinuity and the shocks
-- this is discussed in Sect.~\ref{section ff CWR}).
There is furthermore some discussion
about the distance to the Cyg~OB2 association 
\citep[e.g.,][]{2011A&A...536A..31R}.
A larger distance (1.7 -- 2.0~kpc instead
of our assumed 1.45~kpc) would equally well explain the discrepancy.
Clumping could also be considered as an explanation because it
lowers the mass-loss rates
by a factor 2 -- 3 \citep[e.g.,][]{2006A&A...454..625P}
compared to the smooth wind models used by \citet{2001A&A...369..574V}.
However, the enhanced free-free emission
in a clumped wind will compensate for the lower-mass loss rate, rendering
the clumping explanation unlikely.

Finally, we note that \citet{2012A&A...537A..37M}
provide improved mass-loss rate estimates compared to the
\citet{2001A&A...369..574V} recipe. 
The revised mass-loss rates of the primary and secondary
are a factor $\sim$\,2 lower
and the terminal velocities a factor $\sim$\,1.4 higher than the
ones listed in Table~\ref{table parameters}. This reduces the
predicted fluxes by a factor $\sim$\,4, which leads to values well below
the observed minimum flux. \citeauthor{2012A&A...537A..37M}
however note that their terminal velocities are 35 -- 45 \% too high
compared to observed values.

For the wind momentum ratio
$\eta = \dot{M}_2 \varv_{\infty,2} / (\dot{M}_1 \varv_{\infty,1})$
we find values between 0.97 and 1.76
(using our Table~\ref{table parameters} values).
Most combinations of wind parameters 
give a wind momentum ratio which is $> 1$, i.e. the
secondary star has the stronger wind. Only the O5I+O4III combination gives
a ratio slightly in favour of the primary ($\eta=0.97$).
The corresponding (half) opening angles of the CWR
can be found using Eq.~(3) of
\citet{1993ApJ...402..271E}, giving values between 80\degr\ and 90\degr.

Qualitative constraints on the opening angle can be derived from
the VLBA (Very Large Baseline Array) observation presented by 
\citet[][their Fig.~6]{2006evn..confE..49D}.
This 3.6~cm observation was taken at phase $\sim$\,0.6, i.e. close
to apastron. The CWR shows a clear bow-shape,
curving around an undetected star that is towards the southwest. 
This star must therefore
have the wind with the weaker momentum. At the presumed position
of the weak-wind star, the bow-shaped region shows an indentation
where flux is missing.
At phase 0.6, the stars are separated by 
$\sim\,2700~R_{\sun}$ (projected on the sky).
For the O5I+O3III combination of winds (Table~\ref{table parameters}), 
the surfaces where the optical depth at 3.6~cm is 1
nearly touch, and one can therefore expect some, but not all,
of the non-thermal radio flux to be absorbed.
The VLBA observation therefore seems to
favour an unequal-strength wind scenario. 
In Cyg~OB2\,\#9, a complication arises because the radius where optical depth
is 1 (Table~\ref{table parameters}) is comparable to the separation 
between the two components. In such a case, absorption by the wind in front 
can create bow-shaped emission which is purely an absorption effect and which 
does not relate to the opening angle of the CWR 
\citep[e.g.,][their Fig.~11]{2003A&A...409..217D}.
It is therefore difficult to use the VLBA opening angle to constrain
the momentum ratio of the two stars.
Furthermore, even in an unequal-wind scenario, it is not possible 
to conclude from the VLBA data if it is the primary or the secondary that
has the weaker wind.

\subsection{Free-free contribution colliding-wind region}
\label{section ff CWR}

As the wind-wind collision is adiabatic 
through the majority of the orbit \citepalias{PaperI},
the compressed material in the CWR will be at a high temperature.
It will therefore also contribute to the free-free emission
\citep{2010MNRAS.403.1633P}. Observationally, the high value for the spectral 
index after periastron passage also suggests a thermal (free-free) 
contribution. 
The high temperature of the colliding-wind material
is furthermore attested by the presence of X-ray emission
\citepalias{PaperI}.

\begin{table}
\caption{Free-free flux and spectral index of the simple CWR model
(no synchrotron emission).}
\label{table thermal flux}
\centering
\begin{tabular}{lllllllllllll}
\hline\hline
\multicolumn{1}{c}{model} & \multicolumn{1}{c}{6~cm flux} & \multicolumn{1}{c}{20~cm flux} & \multicolumn{1}{c}{spectral} \\
      & \multicolumn{1}{c}{(mJy)}     & \multicolumn{1}{c}{(mJy)}      & \multicolumn{1}{c}{index}    \\
\hline
                                  no CWR & $ 0.20 -  0.25$ & $ 0.065 -  0.072$ & $ 0.95 -  1.02$ \\
            $T_{\rm CWR} = T_{\rm wind}$ & $ 0.20 -  0.27$ & $ 0.065 -  0.072$ & $ 0.95 -  1.09$ \\
        $T_{\rm CWR} = 2 \times 10^6$\,K & $ 0.21 -  0.28$ & $ 0.065 -  0.076$ & $ 0.97 -  1.13$ \\
        $T_{\rm CWR} = 2 \times 10^7$\,K & $ 0.21 -  0.36$ & $ 0.065 -  0.073$ & $ 0.96 -  1.42$ \\
\hline
\end{tabular}
\tablefoot{The ranges in flux and spectral index
were determined over a range $-0.2$ to $+0.2$ in
orbital phase. 
The CWR size is $3\times$ the separation between the two components
and its thickness is $400~R_{\sun}$. 
The grid extends up to 12\,000~$R_{\sun}$
either side of the origin.
The binary system is assumed to be at a distance of 1.45~kpc.
We used a weak-wind primary and a strong-wind secondary 
in these calculations, corresponding to the O5I+O3III
combination of Table~\ref{table parameters}
($\dot{M}_1=5.66\times 10^{-6}~M_{\sun}\,{\rm yr}^{-1}$,
$\varv_{\infty,1}$ = 2\,079~${\rm km\,s}^{-1}$;
$\dot{M}_2=6.58\times 10^{-6}~M_{\sun}\,{\rm yr}^{-1}$,
$\varv_{\infty,2}$ = 2\,436~${\rm km\,s}^{-1}$).
}
\end{table}

To estimate the free-free emission of the CWR we use the
model from Sect.~\ref{section model}.
For the mass-loss rate and terminal velocity
we use the O5I+O3III combination from 
Table~\ref{table parameters} as this has the largest wind contribution
to the radio flux. 
The size of the CWR is limited to a radius that
is three times the separation between the two components.
We also assign a thickness of $400~R_{\sun}$ to the CWR. 
This value was chosen as it is a significant fraction of
the separation between the two components at periastron
(which is $\sim$\,$500~R_{\sun}$).

In Table~\ref{table thermal flux}, we report the range of flux
values found over the phase $-0.2$ to $+0.2$. We explore
a number of values for the temperature of the CWR ($T_{\rm CWR}$).
As a reference, we also list the value for ``no CWR": in that
model we have a contact discontinuity, where the material from both winds
collides, but the CWR thickness is set
to 0. The values listed for the ``no CWR" model are only about half
of those estimated for the thermal contribution of both stars
in Table~\ref{table parameters}. This is due to the
fact that for each star, all material on the other side of the
contact discontinuity is missing, which is
roughly half of the material (see Sect.~\ref{section winds}).

Table~\ref{table thermal flux} shows there is indeed some
additional flux due to the CWR, which could influence our
interpretation of the Cyg~OB2\,\#9 observations in the low-flux
regime. We do need very high temperatures 
($T_{\rm CWR}=2\times 10^7$~K) over an extended region
to have some detectable influence. This temperature is compatible
with that needed to explain the X-ray observations
\citepalias{PaperI}. As the CWR is adiabatic for most of the orbit,
such high temperature could persist out to large distances from
the CWR apex. At periastron the shocks may become radiative
\citepalias{PaperI}, so the thermal contribution would be less.

At later orbital phases, material with a similar temperature will be present.
Its density will be smaller however. due to the larger separation
between the two stars. This will result in a smaller thermal contribution.
Observationally, this contribution will furthermore
be difficult to disentangle from the much higher non-thermal contribution.

The spectral indexes listed in Table~\ref{table thermal flux} are all
higher than the $+0.6$ nominal value. This is in part a numerical
effect of our simulation: the grid we used covers the 6~cm emitting
region well, but is not large enough to
cover the full extent of the 20~cm emitting region; this leads
to an artificially high spectral index. The relative differences are
significant however, with a higher temperature leading to a higher spectral
index. In a single-star wind, the apparent size increases with wavelength,
leading to the $+0.6$ spectral index. For our model however, we have
a fixed size for the CWR, so the spectral index tends towards the $+2.0$ value
intrinsic to thermal emission (Planck curve). This high
spectral index is in qualitative agreement
with the post-periastron observations.

In summary, a significant contribution of the thermal free-free emission
from the CWR is therefore likely in the low-flux regime around
periastron.

\subsection{Non-thermal emission}

The previous sections showed that the free-free contribution from
both winds, together with free-free emission from
the CWR, can explain at least part of the 
Cyg~OB2\,\#9 flux while it is in the low-flux regime.
The high-flux regime corresponds to a spectral index of 0.2 -- 0.4
(Fig.~\ref{fig lightcurve}), which is more indicative of non-thermal emission.
Indeed, for most of the orbit, Cyg~OB2\,\#9 shows flux values that
are nearly independent of wavelength, indicating a flat
spectral index \citep{2008A&A...483..585V}.  

The observed spectral index
is still larger than the intrinsic one for
non-thermal radiation due to a strong shock, which is expected to be
$-0.5$. A number of effects can change this value. Weaker shocks
would give a more negative index, and therefore cannot explain the
present observations. The inclusion of the Razin effect can substantially
change the spectral index. \citet{2010A&A...519A.111B}
showed that for the shorter-period binary
\object{Cyg~OB2\,\#8A} the Razin effect
can change the index to $+2.0$, thereby simulating a thermal value. After
inclusion of free-free absorption, they found an index of $\sim\,+1.0$
for Cyg~OB2\,\#8A.
The combination of intrinsic synchrotron emission with the Razin effect
and free-free absorption can therefore result in a wide range for the
spectral index. That the
observed values of the spectral index in the high-flux pre-periastron phase
are higher than $-0.5$ is therefore not in contradiction
with non-thermal emission.

There are two phases in the orbit where the projected distance
between the two stars shows a local minimum.
One is at phase 0.96, which coincides with 
the strong drop in flux (at phase $0.934-0.955$). At this phase,
the secondary is closer to us.
Because of the high eccentricity of the binary,
this situation reverses relatively quickly. At phase 0.03, the 
projected distance again has a local minimum, but this time the
primary is closer to us. 

In Sect.~\ref{section winds}
we found that it was most likely that the primary has the weaker wind,
so the synchrotron emission from the CWR should be absorbed less
when the primary is in front (i.e. phase 0.03). 
As this is contradicted by the observations we propose here that
the secondary has the weaker wind. This more easily explains
the observed radio light curve around periastron:
at phase 0.96, the weaker-wind 
secondary is in front of the CWR and blocks some,
but not all,
of the synchrotron emission. At phase 0.03, the stronger-wind
primary is in front and absorbs more of the synchrotron emission
than the secondary did at phase 0.96.

While this seems in contradiction with the results of Sect.~\ref{section winds}.
we should consider the fact that the relation between spectral type --
luminosity class and atmospheric parameters is not unique.
This is clearly shown by 
\citet{2010A&A...524A..98W}, who define spectral-type boxes: these
are regions in the Hertzsprung-Russell
diagram corresponding to a given spectral subtype
and luminosity class. Each spectral-type box has a range in luminosity
and effective temperature. From stellar evolution models they also
assign a range of masses to such a box. The ranges 
for the astrophysical parameters are considerably larger
than those we list in Table~\ref{table parameters}
(e.g. for an O5--5.5 I star, $\log L_{\rm bol}/L_{\sun} = 5.85-6.26$
while our range is only $5.82-5.87$).
If we use this larger range to determine the mass-loss rates and
terminal velocities, we find wind momentum ratios
$\eta = 0.25 - 1.29$.
A primary with the stronger wind is therefore very well possible.

To check in a more quantitative way the hypothesis that a stronger-wind
primary can better explain the observed radio light curve,
we use a model.
A full model would need to include many details, such as the magnetic field,
shock strength, acceleration efficiency and would need to track the electrons
as they move away from the shock 
\citep{2006A&A...446.1001P,2010A&A...519A.111B}.
We postpone such a detailed model to a subsequent paper
(Parkin et al.~2013, in preparation).
Instead we use the simple model from Sect.~\ref{section model}.
We assign two temperatures to the CWR: one is used to represent the
hot, non-relativistic material in the CWR, the other one is the brightness
temperature that represents the synchrotron emission.

\begin{figure}
\centering
\resizebox{9cm}{!}{\includegraphics{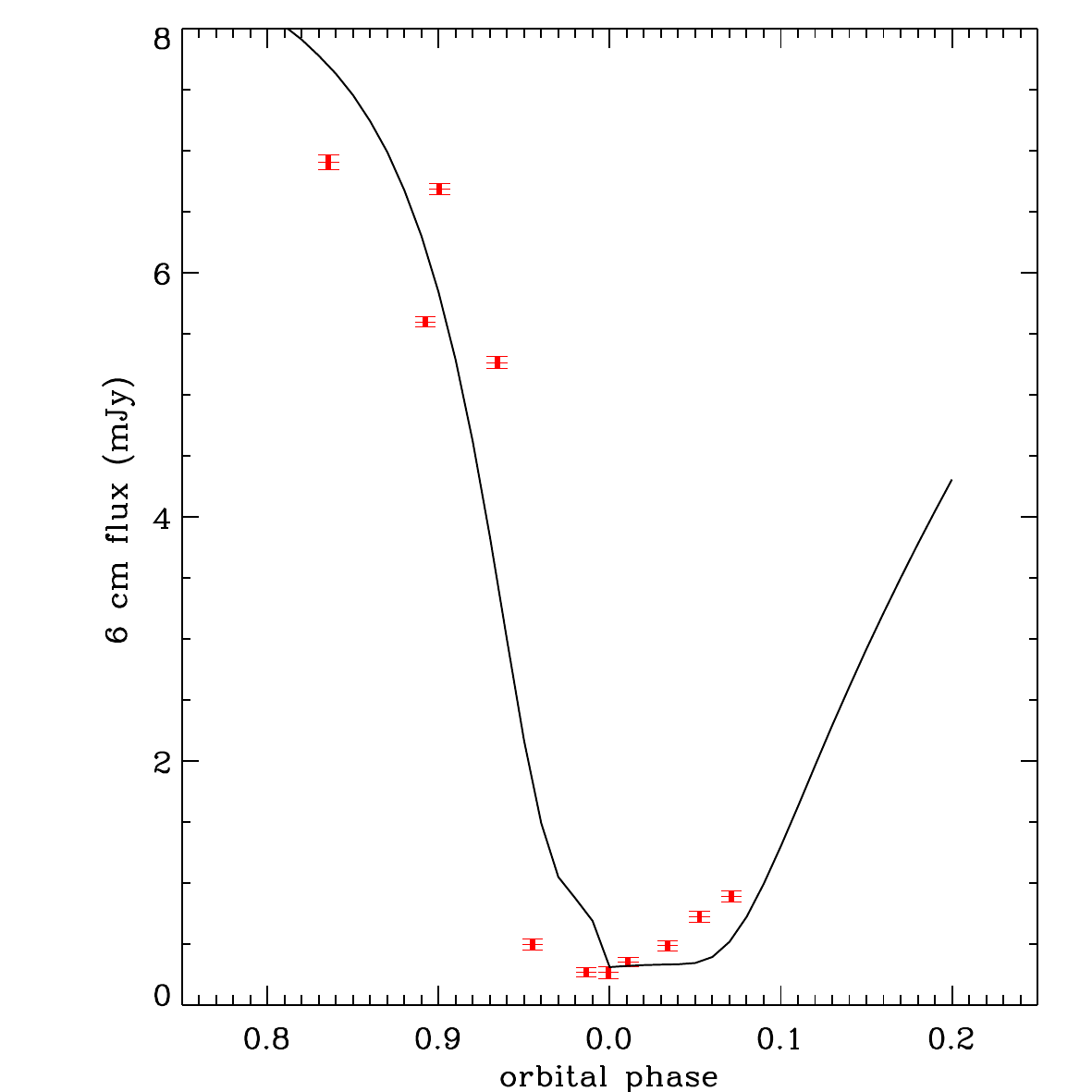}}
\caption{Comparison between the simple CWR model 
that includes synchrotron emission (solid line) and
the observed Cyg~OB2\,\#9 EVLA fluxes 
(symbols as in Fig.~\ref{fig lightcurve}).
The model has a strong-wind primary
($\dot{M}=1.0\times 10^{-5}~M_{\sun}\,{\rm yr}^{-1}$)  and
a weak-wind secondary ($\dot{M}=5.0\times 10^{-6}~M_{\sun}\,{\rm yr}^{-1}$).
Both stars have $\varv_{\infty}$ = 2\,000~${\rm km\,s}^{-1}$.
}
\label{fig fit}
\end{figure}

As it is not the intention of this paper to present a detailed
model
we explore only a small part of the parameter range.
In Fig.~\ref{fig fit}
we show a model that presents an acceptable fit to the 6~cm
data. It is based on a strong-wind primary
($\dot{M}=1.0\times 10^{-5}~M_{\sun}\,{\rm yr}^{-1}$)  and
a weak secondary ($\dot{M}=5.0\times 10^{-6}~M_{\sun}\,{\rm yr}^{-1}$).
These mass-loss rates are roughly the averages from the
\citet{2010A&A...524A..98W} calibration.
Both stars have $\varv_{\infty}$ = 2\,000~${\rm km\,s}^{-1}$.
The size of the CWR is limited to $2\times$ the separation
between the components and its thickness is $200~R_{\sun}$.
The wind temperature is 20\,000~K, the CWR temperature is $10^6$~K
and the brightness temperature of the
synchrotron emitting CWR is $4.0\times 10^8$~K.
We calculate the resulting flux for the orbital phases
covering the range $-0.2$ to $+0.2$ and plot them
on top of the observed Cyg~OB2\,\#9 EVLA fluxes in Fig.~\ref{fig fit}.

Our simple model is able to reproduce the main features of the observed radio 
light curve: the asymmetry between pre- and
post-periastron behaviour, the strong drop in flux around phase 0.955,
the nearly thermal fluxes around periastron
and the slow flux rise after periastron.
The asymmetry between pre- and
post-periastron behaviour and the slow rise after periastron
are direct consequences of our
assumption that the primary has the stronger wind (we cannot reproduce 
the observed asymmetry in the reverse situation). The model does have difficulty
in explaining the more detailed features of the radio light curve.
The drop in the flux around phase 0.955 is not as sharp as observed
and the rise in flux after periastron is slower than observed
(and later becomes faster than the older VLA observations indicate).
We surmise that these are a consequence of the many simplifications
in the model. One relevant effect is our neglect of the Coriolis
force \citep{2008MNRAS.388.1047P}.
Including this will change the shape of the CWR, which could
result in a sharper pre-periastron flux drop and a slower post-periastron
flux rise.

We did not attempt to model the 20~cm fluxes or the spectral index.
The model has no intrinsic calculation of the synchrotron flux or its
spectral index. As we can easily change many parameters
(brightness temperature, CWR size and thickness) to get an acceptable
20~cm light curve, no additional information about the
colliding-wind region would be derived from our modelling of the 20~cm fluxes.

\section{Conclusions}
\label{section conclusions}

As part of a multi-wavelength campaign on the 2011 periastron passage 
of Cyg~OB2\,\#9, we obtained
new 6 and 20~cm radio observations for this highly eccentric
massive O-star binary. 
They show high non-thermal radio fluxes, attributed
to synchrotron radiation emitted by the colliding-wind region
in the pre-periastron phase. 
As the system approaches
periastron the fluxes drop sharply, to levels of free-free emission from 
the stellar winds only. The fluxes then rise again after periastron passage.
These new data
agree very well with the larger set of VLA data presented
by \citet{2008A&A...483..585V}.
The combination of both datasets covers 13
orbits of this system, allowing an accurate determination of the period
($P=860.0 \pm 3.7$ days).

Based on the spectral types of both components, and using theoretical
calibrations \citep{2005A&A...436.1049M, 2001A&A...369..574V}
one would expect the secondary to have the stronger wind 
(i.e. the higher wind momentum, $\dot{M} \varv_{\infty}$).
The calibration of \citet{2010A&A...524A..98W} however, allows for
a larger range of momentum ratios, including those with a stronger-wind
primary.
Using a simple model for the synchrotron emission of the CWR,
we show that a stronger-wind primary
can indeed explain
the main features of the observed radio
light curve: the asymmetry between the pre- and
post-periastron behaviour, the strong drop in flux around phase 0.955,
the nearly thermal fluxes around periastron
and the slow flux rise after periastron.
Additionally, it is likely that the radio fluxes contain some 
free-free contribution from the 
hot and compressed material in the colliding-wind region. 
This free-free contribution may be important
especially in the low-flux regime around periastron passage.

The simple model presented here already 
allows some constraints to be put on 
the parameters of this system. Future, more sophisticated, modelling will 
also include optical, X-ray and interferometric information. It will 
thus provide much better constraints and considerably improve our
understanding of colliding winds in massive star binaries.

\begin{acknowledgements}
We are grateful for the help we received from the NRAO staff 
during the ``Data Reduction Workshop Socorro Spring 2012", where we
learned to apply CASA to our EVLA data.
We thank J. Vandekerckhove for his help with the reduction of the
EVLA data and G. Rauw for comments on an earlier version of the paper.
We thank the referee for his/her comments and for pointing out the
\citeauthor{2010A&A...524A..98W} paper.
D. Volpi acknowledges funding by the Belgian Federal Science Policy Office
(Belspo), under contract MO/33/024.
The Li\`ege team acknowledges support from the European Community's 
Seventh Framework Program (FP7/2007-2013) under grant agreement number 
RG226604 (OPTICON), the Fonds National de la Recherche Scientifique (Belgium), 
the Communaut\'e Fran\c{c}aise de Belgique and the PRODEX XMM and Integral
contracts.
This research has made use of the SIMBAD database, operated at CDS,
Strasbourg, France and NASA's Astrophysics Data System Abstract Service.
\end{acknowledgements}

\bibliographystyle{aa}
\bibliography{radio9}

\end{document}